\def\cm{cm$^{-1}$}
\def\efa{Eu\-Fe$_2$As$_{2}$}

\def\BaFeCoxAs{Ba\-(Fe$_{1-x}$Co$_x$)$_2$As$_{2}$}

\def\bfca{Ba\-(Fe$_{0.92}$Co$_{0.08})_2$As$_{2}$}
\def\bfna{Ba\-(Fe$_{0.95}$Ni$_{0.05})_2$As$_{2}$}

\def\tc{$T_{c}$}
\documentclass[prl,superscriptaddress,twocolumn,showpacs,amsmath,amssymb]{revtex4}
\usepackage[dvips]{graphicx}
\usepackage{graphicx}
\begin{document}
\title{Universal Decomposition of the Low-Frequency Conductivity Spectra of Iron-Pnictides Uncovering Fermi-Liquid Behavior}
\author{D. Wu}
\author{N. Bari\v{s}i\'{c}}
\author{P. Kallina}
\author{A. Faridian}
\author{B. Gorshunov}
\author{N. Drichko}
\affiliation{1.~Physikalisches Institut, Universit\"at Stuttgart,
Pfaffenwaldring 57, 70550 Stuttgart, Germany}
\author{L. J. Li}
\author{X. Lin}
\author{G. H. Cao}
\author{Z. A. Xu}
\affiliation{Department of Physics, Zhejiang University, Hangzhou
310027, People's Republic of China}
\author{N. L. Wang}
\affiliation{
Institute of Physics, Chinese Academy of Sciences, Beijing 100190,
People's Republic of China}
\author{M. Dressel}
\affiliation{1.~Physikalisches Institut, Universit\"at Stuttgart,
Pfaffenwaldring 57, 70550 Stuttgart, Germany}
\date{\today}
\begin{abstract}
Infrared reflectivity measurements on 122 iron-pnictides reveal
the existence of two electronic subsystems. The one gapped due to
the spin-density-wave transition in the parent materials, such as
\efa, is responsible for superconductivity in the doped compounds,
like \bfca\ and \bfna. Analyzing the dc resistivity and
scattering rate of this contribution, a hidden $T^2$ dependence is
found, indicating that superconductivity evolves out of a
Fermi-liquid state. The second subsystem gives rise to incoherent
background, present in all 122 compounds, which is basically
temperature independent, but affected by the superconducting
transition.
\end{abstract}
 \maketitle
In the course of the past year, it became apparent that both the
normal and superconducting states of iron-pnictides are more
complex than in conventional metals \cite{Norman08,Ishida09}.
Understanding their magnetic and structural orders in parallel to
superconductivity is one of the main challenges. An additional
complication is the pre\-sence of several bands close to the Fermi
energy where nesting might be crucial for the magnetic order
\cite{Dong08,Singh08}.
Sub\-stituting Ba by K in BaFe$_2$As$_2$, for instance, introduces
holes and affects structural and spin density wave (SDW)
transitions, which are both suppressed for the benefit of
superconductivity \cite{Rotter09}. When Co or Ni partially
replaces Fe also both transitions are suppressed and
superconductivity is observed up to $T_c=25$~K
\cite{Sefat08,Li09}.
The proximity of a magnetic ground state raises the question about
the role of magnetic fluctuations \cite{Ning09a}, which
--~together with the observed anomalous transport behavior~--
evokes a quantum-phase-transition scenario for superconductivity
\cite{Chu09,Gooch09}.
Here we compare the optical properties on different
122-iron-pnictide compounds and find that the low-frequency
conductivity spectra can be universally decomposed into a broad
temperature independent part and a narrow zero-frequency
contribution. The latter one exhibits a Fermi-liquid behavior and
governs the physical behavior of these materials.

Single crystals of \efa, \bfca, and \bfna\ were grown using FeAs
as self-flux dopants \cite{Chu09,Li09} and characterized by X-ray,
EDX-microanalysis, transport and susceptibility measurements
[Fig.~\ref{fig:Fig3}(a,b)]. The high reproducibility of the
resistivity measurements and the sharp transition observed in the
susceptibility indicate that the samples are homogeneously doped
\cite{Barisic08}. The temperature dependent optical reflectivity
(in $ab$ plane) was measured in a wide frequency range from 20 to
37\,000~\cm\ using a series of spectrometers. The low-frequency
extra\-polation was done according to the dc conductivity measured
on the same crystals. The complex optical conductivity
$\hat{\sigma}(\omega)=\sigma_1(\omega)+{\rm i}\sigma_2(\omega)$
was calculated from the reflectivity spectra using Kramers-Kronig
analysis.
\begin{figure}[b]
\centering
\includegraphics[width=0.85\columnwidth]{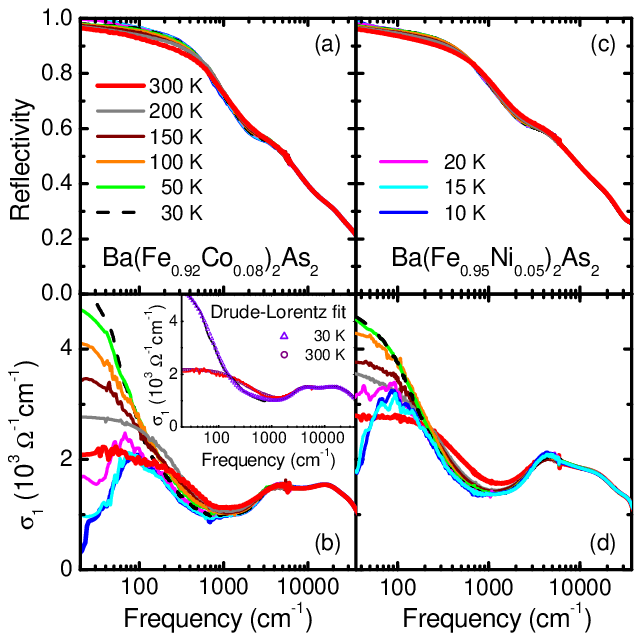}
\caption{\label{fig:Fig2} (Color online)
Optical reflectivity and conductivity of \bfca\ and \bfna\
measured at different temperatures between 10 K and 300~K in a
wide spectral range. Examples of Drude-Lorentz fits are displayed
in the inset for two temperatures $T=30$~K and 300~K.}
\end{figure}

In Fig.~\ref{fig:Fig2} we present the in-plane optical properties
of \bfca\ and \bfna\ at different temperatures. The reflectivity
is metallic but does not exhibit a clear-cut plasma edge similar
to other 122 iron-pnictide compounds
\cite{Hu08,Li08,Nakajima09,Chen09,Hu09c,Yang09,Wu09}.
\begin{figure*}
\centering
\includegraphics[width=14.8cm]{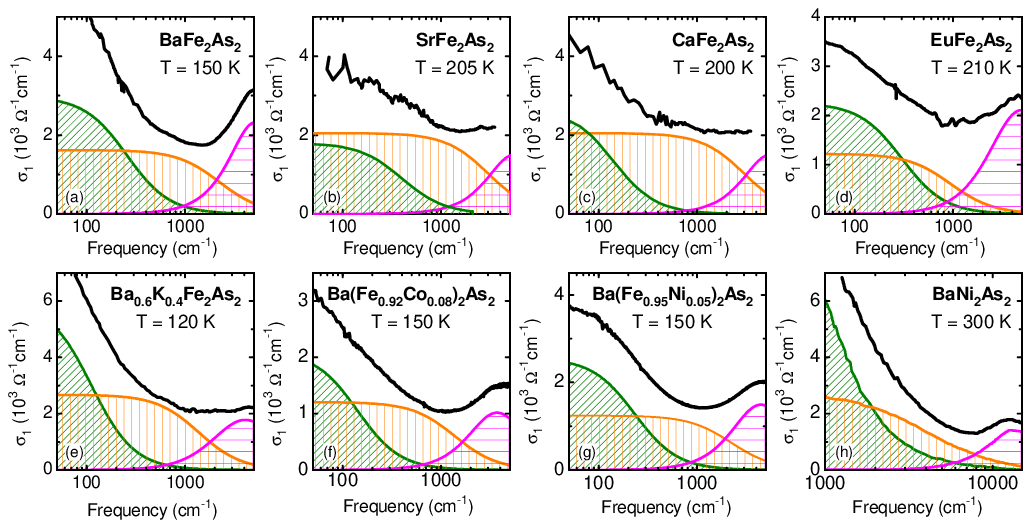}
\caption{\label{fig:Fig1} (Color online)
In the metallic state the optical conductivity of different
iron-pnictides can always be described by two Drude terms
($\sigma_N$-densely shaded green and $\sigma_B$-medium shaded
orange) and an oscillator in the mid-infrared (sparsely hatched
magenta). The spectra of the parent compounds at temperatures
above the SDW transition: (a)~BaFe$_2$As$_2$ \cite{Hu08},
(b)~SrFe$_2$As$_2$ \cite{Hu08}, (c)~CaFe$_2$As$_2$
\cite{Nakajima09}, and (d)~\efa\ \cite{Wu09}. For $T>T_c$ the
conductivity spectra of the electron doped superconductor
Ba$_{0.6}$K$_{0.4}$Fe$_2$As$_2$ \cite{Li08}, the hole doped
superconductors (f)~\bfca\ and (g)~\bfna\ are very similar.
(h)~When Fe is completely substituted by Ni, the material is more
metallic, but the room-temperature spectrum of BaNi$_2$As$_2$
\cite{Chen09} can be decomposed in the same way with a larger
spectral weight of the Drude components.}
\end{figure*}
The optical conductivity contains a broad double-peak in the mid-
and near-infrared that does not change considerably with
temperature. As $T$ decreases, a Drude-like contribution develops
and becomes narrower. Both features are separated by a minimum in
$\sigma_1(\omega)$ around 1000~\cm. The parent compound \efa\ was
discussed in Ref.~\onlinecite{Wu09}.

When we try to fit the normal-state spectra by a minimum number of
Drude and Lorentz contributions \cite{DresselGruner02}, we obtain
a satisfactory description only by using two Drude terms
\cite{remark5}. Importantly, this decomposition holds universally
for different 122-iron-pnictides. This is illustrated in
Fig.~\ref{fig:Fig1} where the optical conductivity is displayed in
a large frequency range for parent compounds $X$Fe$_2$As$_2$ ($X$
= Ba, Sr, Ca, Eu), as well as for hole-doped
Ba$_{0.6}$K$_{0.4}$Fe$_2$As$_2$ and for the completely substituted
BaNi$_2$As$_2$ \cite{Hu08,Nakajima09,Wu09,Li08,Chen09}. A simple
oscillator models the mid-infrared properties sufficiently well. A
broad contribution $\sigma_B$ mimics the considerable background
conductivity that seems to be similar for all compounds and does
not change appreciably with temperature. As will be discussed
later, the distinct properties of the particular material are
solely determined by the second Drude term $\sigma_N$ with a width
of approximately 100~\cm\ in the case of \bfca. Its spectral
weight is smaller by a factor of 3-5 compared to $\sigma_B$. Upon
cooling, the spectral weight is
roughly conserved for each of these two
subsystems.
Our finding is in accord with angle-resolved photo\-emission (ARPES) data \cite{Evtushinsky09} which
indicate that the multi-band structure does not change considerably upon
doping and temperature.

In order to analyze the temperature dependence of $\sigma_N$ at
$\omega=0$, we have measured the dc resistivity [see Fig.~\ref{fig:Fig3}(a)], then simply deduce a constant $\sigma_B$ from the graphs
displayed in Fig.~\ref{fig:Fig1} and consider $\sigma_N(T) =
\sigma_{\rm total}(T) - \sigma_B$. The result is surprising but
very simple, namely $\rho_N = 1/\sigma_N\propto T^2$ over the broad temperature range, as shown in Fig.~\ref{fig:Fig3}(c). Such a temperature dependence is a clear sign of Fermi liquid behavior. Importantly, an independent confirmation of this finding is obtained directly from the decomposition of the low-frequency
conductivity of Ba(Fe$_{1-x}M_x)_2$As$_2$. The Drude fit $\sigma_N(\omega)= \sigma_{N}(\omega=0)/(1+\omega^2\tau_N^2)$
yields a scatting rate $1/\tau_N(T)\propto T^2$ as well
[Fig.~\ref{fig:Fig3}(d)]. These results evidence a deep physical consistency of our decomposition.

We want to point out that due to the broad frequency range covered
by only two Drude contributions, the fits are robust and give well
defined parameters $\sigma_B$, $\sigma_N(T)$, and $1/\tau_N(T)$.
Furthermore, we restrain ourselves by assuming $\sigma_B$ basically as
temperature independent; and since the dc-conductivity is
determined separately, that leaves only $1/\tau_N(T)$ as a
fit parameter. Moreover, as
a result of such fit procedure it turns out that the coherent
Drude term preserves its spectral weight
within an uncertainty of 2\%.
The fact that the plasma frequency does not change much with temperature
reinforces the conclusion that the coherent
Drude describes a classical Fermi liquid.
\begin{figure}
\centering
\includegraphics[width=0.85\columnwidth]{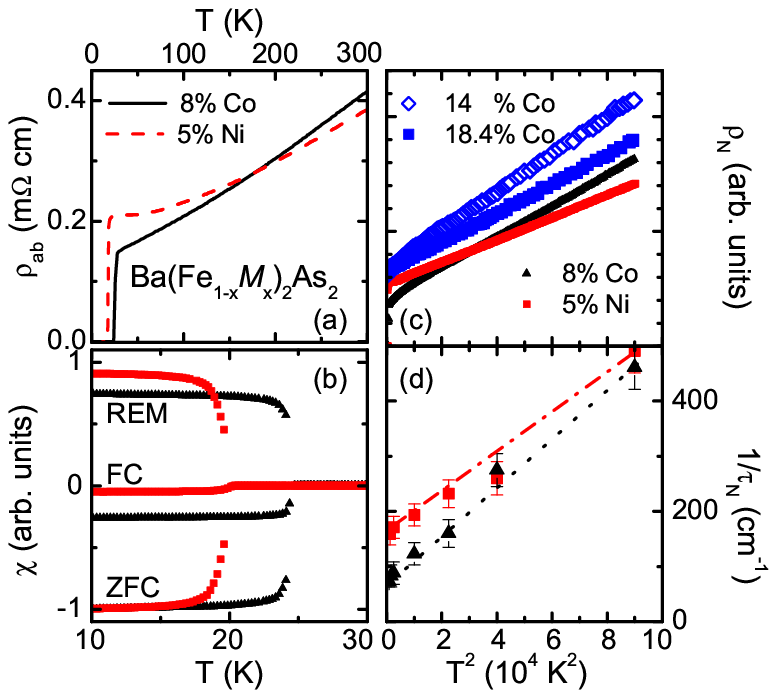}
\caption{\label{fig:Fig3} (Color online)
(a) Temperature dependence of the electrical resistivity of \bfca\
and \bfna\ crystals. (b) The field-cooled (FC), zero-field-cooled
(ZFC) and remanent (REM) susceptibility exhibit a sharp change at
the transition temperature ($T_c=25$~K and 20~K). (c)~``narrow
Drude'' contribution to dc resistivity plotted as a function of
$T^2$, as obtained by decomposition $\rho_N(T) = [\sigma_{\rm
total}(T) -\sigma_B]^{-1}$, where the $\sigma_B$ is constant in
temperature. The data is presented in arbitrary units since Ba(Fe$_{0.86}$Co$_{0.14}$)$_2$As$_2$ and
Ba(Fe$_{0.816}$Co$_{0.184}$)$_2$As$_2$ are taken from
Ref.~\onlinecite{Chu09} where no absolute values are given. (d)
Temperature dependence of the scatting rate $1/\tau_N(T)$ obtained
from the fit of the low-frequency optical data by the Drude
model.}
\end{figure}

Our findings are in agreement with the
conclusion drawn from a comprehensive Hall-effect study on the
series \BaFeCoxAs\ \cite{Rullier09} indicating that the
iron-pnictides are good Fermi liquids. As discussed there the
values of scattering rate and the $T^2$ behavior agree reasonably
with the weak coupling limit under assumption that the effective
Umklapp scattering is comparable to the Fermi energy. Furthermore
other iron-pnictides behave in a very similar way, as demonstrated
on \bfna\ in Fig.~\ref{fig:Fig3}(c), confirming our general view.
Notably, a $T^2$ behavior of the resistivity is also found in 1111
compounds \cite{Sefat08b,Suzuki09}.

We note that our discovery of a hidden Fermi-liquid behavior of the governing
charge carriers is at variance with previously reported findings.
Those works analyzed the resistivity prior to its decomposition,
having both contributions, $\sigma_N(T)$ and $\sigma_B$, mixed.
The ``anomalous'' temperature dependence of the resistivity
($\rho_0 + AT^n$ or $\rho_0 + AT + BT^2$) and the proximity of the
magnetic phase have been considered as an indication that the
superconducting ground state is a consequence of underlying
quantum criticality \cite{Chu09,Gooch09,remark5}.
Indeed as shown in Fig.~\ref{fig:Fig3}(a) the total resistivity
is clearly not quadratic in temperature and can be well fitted with $AT^n$
with $n= 1.25$ for \bfca\ and $n=1.5$ for \bfna. Only when the
incoherent part, determined by optical conductivity, is
subtracted, the $T^2$ law appears in both samples, as demonstrated
in Fig.~\ref{fig:Fig3}(c).
With doping the dc resistivity gets considerably smaller
\cite{Rullier09,Kasahara09}, because $\sigma_n$ grows.
Consequently the contribution of the incoherent term to $\sigma_{\rm total}$
becomes less important. This explains the change in the overall power-law coefficient upon doping \cite{Kasahara09}, eventually leading to $n=2$. Our decomposition of the optical conductivity and dc resistivity holds for all analyzed 122 iron-pnictides and reveals the Fermi liquid behavior in $\sigma_n$. This result is universal and physically transparent \cite{remark5}.
\begin{figure}
\centering
\includegraphics[width=.85\columnwidth]{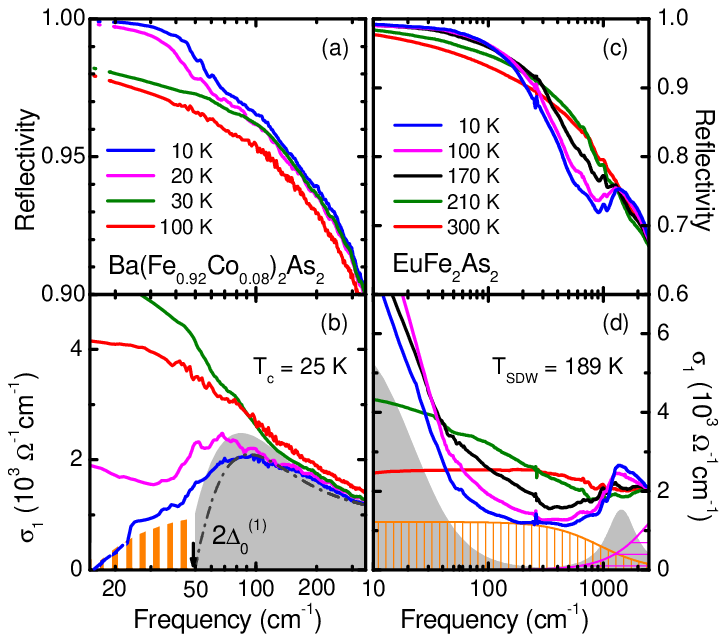}
\caption{\label{fig:Fig4} (Color online)
Reflectivity and conductivity of \bfca\ and \efa\ above and below
the superconducting and SDW transitions.
(a) The way the reflectivity of \bfca\
approaches unity is indicative for the development of the
superconducting gap. The dotted lines corresponds to the
extrapolations. (b) The optical conductivity $\sigma_1(\omega,T)$
is gradually depleted up to approximately 150~\cm. The dash-dotted
curve is calculated for the lowest
temperature assuming one gap $2\Delta_0^{(1)}=50$ ~\cm and the shaded area by assuming gaps $2\Delta_0^{(1)}=50$~\cm\ in the
$\sigma_N(\omega)$ component (mainly grey area) and
$2\Delta_0^{(2)}=17$~\cm\ in the $\sigma_B(\omega)$ background
(mainly orange area).
(c,d) For $T<T_{\rm SDW}$, a gap in the excitation spectrum of \efa\ opens. Part of the spectral weight of $\sigma_N$ shifts from low-frequencies to
above approximately 1000~\cm; but a small Drude-like peak remains.
The incoherent background $\sigma_B(\omega)$ remains basically
temperature independent. The fit by the Drude-Lorentz model
corresponds to $T=10$~K. }
\end{figure}

Next we analyze the superconducting state in detail. Upon passing
the superconducting transition the reflectivity of \bfca\
increases considerably below 70~\cm, as can be seen in
Fig.~\ref{fig:Fig4}(a); for \bfna\ a similar behavior is observed
around 60~\cm. The way it approaches unity with a change in
curvature causes a gap-like structure in the optical conductivity
below 100~\cm\ [Fig.~\ref{fig:Fig4}(b)]. The removal of spectral
weight in $\sigma_1(\omega,T)$ compared to the normal state
conductivity $\sigma^{(n)}(\omega,T\approx 30~{\rm K})$ extends
all the way up to approximately 150~\cm.
From the missing area \cite{DresselGruner02}, $A=\int\left[\sigma^{(n)}_1(\omega)-
\sigma^{(s)}_1(\omega)\right]{\rm d}\omega$ we estimate the penetration depth
to $\lambda=c/\sqrt{8A}=(3500\pm 350)$~\AA\ and $(3000\pm 300)$~\AA\ for
the Co and Ni compounds, respectively.
Notably, both materials obey the universal scaling relation of the superfluid
density $\rho_s\propto A$ and the transition temperature $T_c$
\cite{Homes04,Wu10}.
This indicates that the main features of the superconducting state are well captured by the optical conductivity.

To get further insight, we model our spectra by the
BCS-theory for the frequency and temperature dependent
conductivity \cite{Mattis58}. As seen by the dash-dotted lines in
Fig.~\ref{fig:Fig4}(b), a decent description is obtained with gap
value of 50~\cm.
Following the suggestions of two isotropic gaps
\cite{Evtushinsky09,Khasanov09,Mazin08}, both opening
simultaneously below \tc, we obtain an even better description
with $2\Delta_0^{(1)}=50$~\cm\ in the narrow Drude term only, and
a second smaller gap $2\Delta_0^{(2)}=17$~\cm\ which affects the
broad background \cite{remark1}. Similarly, the description of Ba$_{0.6}$K$_{0.4}$Fe$_2$As$_2$ \cite{Hu09c} suggests two superconducting gaps. This indicates that the hole-doped iron-pnictides behave in the same way as the electron-doped compounds and that our conclusions are universal in this sense.

The gap values obtained from our study are $2\Delta_0^{(1)}/k_BT_c=2.5 -3$, and 1 for the small gap. ARPES experiments on the K doped compound yield to higher gap values of $2\Delta_0^{(1)}/k_BT_c=6.8$ and $2\Delta_0^{(2)}/k_BT_c<3$ \cite{Evtushinsky09}. It is worthwhile to recall the situation of
the two-gap superconductor MgB$_2$, for which several methods
evidence two gaps, while far-infrared measurements see only one.
By now it is not understood why the optical results agree better with the
smaller gap value, although they should be sensitive only to the
large one \cite{Kuzmenko07}.
Due to the proximity of the superconductivity and magnetic ground state it is instructive to compare our results with those obtained on parent compounds. Upon passing through the SDW transition of \efa, for instance \cite{Wu09}, a
gap in the excitation spectrum opens since parts of the Fermi
surface are removed. Importantly, as demonstrated in
Fig.~\ref{fig:Fig4}(d), only spectral weight of $\sigma_N$ shifts
from low-frequencies to above approximately 1000~\cm. The
incoherent background $\sigma_B$ remains basically
unchanged. A small Drude-like peak (accounting for the metallic dc
properties) remains and becomes very narrow as $T$ is reduced. On the other hand, the metallic state of the
Ba(Fe$_{1-x}M_x)_2$As$_2$ compounds ($M$ = Co or Ni, close to
optimally doping) does not exhibit any trace of the SDW pseudogap
(Fig.~\ref{fig:Fig2}). Upon cooling down to the superconducting
transition, $\sigma_N (\omega)$ gradually narrows and grows
[Fig.~\ref{fig:Fig2}(b,d)]. Importantly, at $T_c$ a dominant superconducting gap develops in $\sigma_N$, i.e. it is associate with the same parts of the Fermi surface gaped by the SDW in parent compounds.
Our analysis of the optical conductivity in the normal and broken-symmetry ground states of iron-pnictides, reveals two sorts of conduction electrons associated with $\sigma_B$ and $\sigma_N$ (Fig.~\ref{fig:Fig1}). The broad term $\sigma_B$ is temperature independent. We ascribe it to an incoherent subsystem which seems to be a common fact in all iron-pnictides. The narrow Drude-like contribution $\sigma_N$ dominates the conduction in the pnictides. It reveals a $T^2$ behavior in the resistivity and scattering rate, inferring that superconductivity grows out of a Fermi liquid. Furthermore the dominant gap at $2\Delta^{(1)}$ in electron doped Ba(Fe$_{1-x}M_x)_2$As$_2$ and the SDW gap in parent compound \efa\ develop in $\sigma_N$, which signifies that the underlying ground states of iron-pnictides involve the same parts of the Fermi surface.

We acknowledge the help of J. Braun and E.S. Zhukova. N.B.
is supported by the Alexander von Humboldt Foundation.
N.D. thanks the Magarete-von-Wrangell-Pro\-gramm. The is supported
by NSF of China.


\begin{thebibliography}{99}
\bibitem{remark5}
Other decompositions of the optical conductivity than suggested
here might be tempting, too. For example, the next simplest one is
to fit the low-frequency conductivity by a Drude and a Lorentzian
terms. However, there is not really any physical meaning of the
Lorentzian, and furthermore its position, width and amplitude
strongly change with temperature which is highly anomalous.
\bibitem{remark1}
The single-gap model significantly deviates from the data in the range of
high accuray between 100 and 300~\cm, because the incoherent
$\sigma_B$ has a large width.
The hatched area in Fig.~\ref{fig:Fig4}(b) indicates
the superior description by the two-gap BCS model; it does not
imply an unambiguous proof of its size. The
experimental results for \bfna\ are qualitatively identical, but
with a lower value $2\Delta_0^{(1)}=35$~\cm\ in the narrow Drude
term. The small gaps fall below the range in which reliable data
could be acquired.
\end{thebibliography}
\end{document}